\documentclass{aa}
\usepackage{graphicx,natbib}
\bibpunct{(}{)}{;}{a}{}{,}
\begin{document}
   \title{An H$\alpha$ survey aiming at the detection of extraplanar diffuse 
ionized gas in halos of edge--on spiral galaxies
\thanks{Based on observations collected at the European Southern Observatory, 
Chile (ESO No.~63.N--0070, ESO No.~64.N--0034, ESO No.~65.N.--0002).}}

   \subtitle{I. How common are gaseous halos among non--starburst galaxies?}

   \author{J. Rossa \thanks{Visiting Astronomer, German--Spanish Astronomical 
Centre, Calar Alto, operated by the Max--Planck--Institute for Astronomy, 
Heidelberg, jointly with the Spanish National Commission for Astronomy.}
          \inst{1,2}
          \and
          R.--J. Dettmar\inst{1}
          }

   \offprints{J.~Rossa (jrossa@stsci.edu)}

   \institute{Astronomisches Institut, Ruhr--Universit\"at Bochum,
              D--44780 Bochum, Germany\\
              \email{jrossa@stsci.edu, 
              dettmar@astro.ruhr-uni-bochum.de}
         \and
             Space Telescope Science Institute, 3700 San Martin Drive,  
             Baltimore, MD 21218, U.S.A. (present address)}

   \date{Received 28 January 2003 / Accepted 15 April 2003}

   \abstract{
In a series of two papers we present results of a new H$\alpha$ imaging 
survey, aiming at the detection of extraplanar diffuse ionized gas in 
halos of late--type spiral galaxies. We have investigated a sample of 74 
nearby edge--on spirals, covering the northern and southern hemisphere. In 
30 galaxies we detected extraplanar diffuse emission at mean distances of 
$|z|\sim1-2$\,kpc. Individual filaments can be traced out to $|z|\leq6$\,kpc 
in a few cases. We find a good correlation between the FIR flux ratio 
($S_{60}/S_{100}$) and the SFR per unit area ($L_{\rm FIR}/D^2_{25}$), based 
on the detections/non--detections. This is actually valid for starburst, 
normal and for quiescent galaxies. A minimal SFR per unit area for the lowest 
$S_{60}/S_{100}$ values, at which extended emission has been detected, was 
derived, which amounts to $\dot{E}_{\rm A25}^{\rm{thres}} = 
(3.2\pm0.5)\times10^{40}\rm{\,erg\,s^{-1}\,kpc^{-2}}$. There are galaxies 
where extraplanar emission was detected at smaller values of 
$L_{\rm FIR}/D^2_{25}$, however, only in combination with a significantly 
enhanced dust temperature. The results corroborate the general view that 
the gaseous halos are a direct consequence of SF activity in the underlying 
galactic disk.
   }

\titlerunning{An H$\alpha$ survey aiming at the detection of eDIG in edge--on 
spiral galaxies}
   \maketitle
%

\section{Introduction}

In recent years observations revealed the presence of an extraplanar diffuse 
ionized gas (eDIG) layer \citep[e.g.,\,][]{Ra96,Ro00} in a few late--type 
spiral galaxies, similar to the {\em Reynolds layer} in the Milky Way 
\citep[e.g.,\,][]{Re99}. The occurrence of such a layer in a galaxy is most 
likely a direct consequence of the strength of SF activity in the disk of 
each individual galaxy. The strength of SF activity both on local and global 
scales plays an important role. It was already assumed in previous studies 
that the star formation activity in the disk of spiral galaxies is correlated 
with the presence of DIG and its extent in the halos of these galaxies.     

Current models, trying to explain the gas and energy transport from the disk
into the halo, have been developed over a decade ago \citep[e.g.,\,][]{No89} 
where the gas from the disk emanates into the halo, the so called {\em 
chimney} phenomenon, driven by collective SNe. A few theoretical concepts 
for outflow phenomena besides chimneys have been established as well, such 
as {\em galactic fountains} \citep[e.g.,\,][]{Sh76,Br80}, or the superbubble 
outbreak scenario \citep{Ma88}. Observations have confirmed a disk--halo 
connection. This was successfully demonstrated for several edge--on spirals 
\citep[e.g.,\,][]{De90,Ra90,Pi94,Ra96,Ro00}.  

The models can explain some of the observed outflow phenomena of the 
interstellar gas, but they are far away from giving a coherent picture
of the whole situation. As it is argued by \citet{Le96} many starburst 
galaxies follow the condition of a superwind, which drives the outbraking 
gas from bubbles into the halo. The theory of such a superwind was developed 
in the mid eighties \citep{Ch85}. Recent studies of actively star--forming 
regions in edge--on galaxies gave evidence for a superwind \citep[for a 
recent review see][]{He01}, such as in the case of NGC\,4666 \citep{Da97}. 
Similar conditions apply for M\,82 and NGC\,4631 \citep{Bl88}, and for 
NGC\,3079 \citep{Ce02}. It is, however, currently unclear whether it is also 
a superwind that causes outflows in galaxies with much lower SF activity. 

Most of the recent studies focused on the different morphology of the DIG
(e.g.,\,plumes, filaments, loops) in order to describe the overall
distribution and structure in halos, for example the spectacular filaments 
of the nearby galaxy NGC\,55 \citep{Fe96}. Only one recent study addressed
the correlation with global properties from a larger sample of objects.
However, this study \citep[][hereafter LH95]{Le95} selected only measures 
such as the IR--luminosity as the primary selection criterion, but this can 
be biased by strong local star--forming activity. Therefore, their sample is 
strongly biased {\em against} low luminosity galaxies which are not covered 
at all. \looseness=-2    

It is therefore important to consider selection criteria in addition to the
IR properties. To study an unbiased sample (in the sense of FIR emission), it
is essential to deal with a distribution of IR--luminosities, containing
galaxies with {\em low} \underline{and} {\em high} values of the IR--fluxes,
rather than selecting only IR bright galaxies, in order to find the division
between those galaxies showing outflows, and those which do not bear any
signs of disk--halo interaction at all. Furthermore, it could be important to
make use of a sample which includes also different galaxy types in the sense
of evolution, covering early--type spirals (Sa) as well as late--type spirals
(Sc). Since there are indications for DIG in early--type spirals
\citep[cf.\,][]{De92}, this aspect has to be investigated, too. Indeed,
there were a few cases where strong star formation has been found among some
early--type face--on spirals \citep[e.g.,\,][]{Ha99}. 

In an approach to study the star--forming activity, which is traced by the
diffuse ionized gas (DIG), we have defined specific selection criteria
concerning the sample of our program galaxies. For this reason we have 
observed a large sample of galaxies, comprising edge--on spiral galaxies 
with very different FIR luminosities and different Hubble types. Since the 
SFR is correlated with the FIR luminosity, we will try to trace the fainter 
end of the FIR luminosity distribution, to find the division between galaxies 
with high star formation activity (e.g.,\,starbursts) and galaxies with low 
star formation activity (non--starbursts) via the 
{\em detection/non--detection} of star--formation--driven outflows. By clearly
showing non--detections (at the level of our sensitivity limit) of any DIG
features in the less active galaxies of this sample, we will be able to
derive the minimal SFR per unit area which is necessary to lead to such
outflows. This is important for the understanding of the physics of the
interstellar medium and for the evolution of galaxies, since this process is
closely related to mass loss and chemical evolution. Finally, these
observations can be compared with recent theoretical models, such as the
superbubble outbreak scenario \citep{Ma99}.  

Some of the basic questions we want to address in this investigation are
the following: How common are galactic gaseous halos and what is the 
morphology of eDIG? What properties (parameters) determine the presence of 
galactic outflows? Is there a minimal SFR per unit area, at which gaseous 
outflows can be detected?


\section{H$\alpha$ survey selection criteria}

The diffuse ionized gas (DIG), frequently also called WIM (warm ionized
medium) is best traced by the relatively bright H$\alpha$ emission. From a
first sub--sample, which was discussed in detail in \citet{Ro00}, it became
clear that there is a need for a larger sample in order to quantify the
observed outflows (i.e. to derive a measure for the minimal SF activity per
unit area), and to have a large sample for good statistical information as
well. Whereas FIR bright galaxies have already been studied in detail in the
context of disk--halo interaction \citep{Le95,Le96}, there was a need for 
studying a broader distribution in FIR flux coverage. The target galaxies 
for our unbiased (in the sense of SF activity) H$\alpha$ survey have been 
selected primarily by the broad coverage of the FIR luminosity (for those 
with high \underline{and} those with low FIR luminosities) and the edge--on 
character of the galaxies, where the halo separates from the disk very well 
in projection. Moreover, the selected galaxies should not be too large, in 
order to fit into the field of view (FOV) of the used instruments and to 
avoid assembling a mosaic from several sub--frames, which would be time 
consuming. This constraint applies to almost all of our target galaxies. On 
the other hand the galaxies should not be too small, so that the possible 
DIG features cannot be resolved. Therefore, a minimum apparent diameter of 
$3'$ was chosen, to make an useful investigation. To study any DIG features
in the halos, the galaxies should be almost edge--on. We chose a minimum
inclination angle of $\geq76^\circ$, determined by the galaxy axial
ratios (see below). The selection of the final target galaxies was based on
the RC3 catalog \citep{Va91}, and also on the {\em Flat Galaxy Catalogue} 
(FGC), published by \citet{Ka93}. To see, whether or in how far companion 
galaxies may trigger active star formation, we have included in our sample 
some edge--on galaxies with associated companions (also taken from the 
above mentioned catalogs), to study this aspect, too. The observations for 
this survey have been divided into three individual observing runs on the 
southern hemisphere, and two on the northern hemisphere, to cover the 
complete R.A. interval and to have an even more homogeneous sample.        

The final selection criteria for our H$\alpha$ survey are shortly summarized
below 

\vspace{0.2cm}

\begin{itemize}
\item spiral galaxies 
\item unbiased galaxy sample with respect to the FIR flux
\item broad coverage of SF activity
\item $i \geq 76^\circ$ (majority has $i \geq 80^\circ$)
\item $3' \leq D_{25} \leq 12'$
\item $v_{\rm rad} \leq 6000\,{\rm km\,s^{-1}}$
\item availability of the desired H$\alpha$ filters according to redshifts of
the galaxies
\end{itemize}        

\vspace{0.2cm}

The observations for the northern hemisphere galaxies have been performed 
with CAFOS\footnote{Calar Alto Faint Object Spectrograph} in direct imaging 
mode, attached to the CAHA 2.2\,m telescope (Spain). The southern hemisphere 
galaxies have been observed with DFOSC\footnote{Danish Faint Object 
Spectrograph Camera} in imaging mode, attached to the Danish 1.54\,m 
telescope at La Silla/Chile. More observational details are presented in our 
second paper, which gives the full details on each individual galaxy 
\citep{RoDe01}.

All the galaxies, which did fit the above mentioned selection criteria, were 
first inspected visually on the basis of the Digitized Sky Survey (DSS).
Initially, about 200 galaxies were regarded as suitable candidates for
our investigation. However, further constraints did arise. In several cases
bright foreground stars, that were projected onto the galaxy disk, have been
detected on the DSS images. Those galaxies have been rejected, since bright
stars will cause artifacts which make it difficult to locate DIG. 
Furthermore, galaxies in compact groups have been rejected as well, since
close galaxy pairs may also trigger outflows through interactions. However,
we did include wide pairs of galaxies in a few cases. A certain fraction of
the galaxies that were in our initial list have already been observed by other
investigators \citep[e.g.,\,][]{Pi94,Le95,Ra96,Ho99}. Except in a very few 
cases (for comparison with our sensitivities) we did refrain from 
re--observing these galaxies. Moreover, several galaxies have been dropped 
from our list, since their redshifts were in conflict with the wavelength 
coverage of the available H$\alpha$ filters. 
  
Finally, about 120 galaxies did fulfill all of our selection criteria. We 
observed 65 of these galaxies, given the object visibility at our various
scheduled observing runs. Adding the nine observed galaxies from our recent 
study \citep{Ro00}, our survey consists of 74 galaxies in total. Although 
this is the largest H$\alpha$ survey to date investigating DIG in edge--on 
galaxies, it should be mentioned that it is not a statistically complete 
sample. Nevertheless, we have carefully selected our galaxies, to study the 
fainter end of the FIR luminosity distribution. In fact, 12 of our galaxies 
have not been detected with IRAS in the FIR. We therefore do have a 
sufficiently high fraction of galaxies, with very low FIR--fluxes, which 
populate the faint end of the FIR--distribution, in which we were interested 
in. 


\section{Results}

In the following we present some important properties of the survey 
galaxies. In Fig.~1 we have plotted a histogram showing the distribution of 
the galaxy types of our survey galaxies. It is evident from this plot that 
the early--type spirals are underrepresented, as we gave more weight to the 
late--type spirals. 

\vspace{0.2cm}
\begin{figure}[h]
\begin{center}
\hspace{0.1cm}
\rotatebox{270}{\resizebox{5cm}{!}{\includegraphics[bb=28 60 490 773,clip]{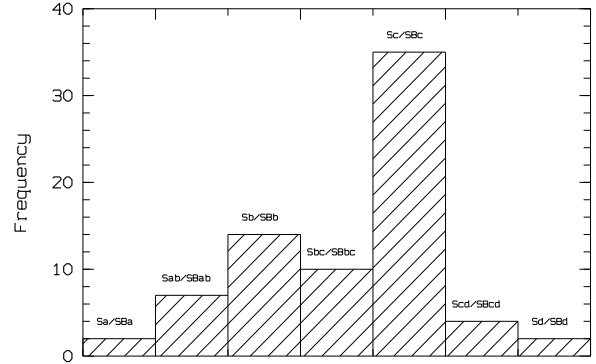}}}
\caption[]{\small Histogram showing the distribution of the galaxy type for 
the 74 survey galaxies.}
\end{center}
\end{figure}
\vspace{0.2cm}     

The distribution of the radial velocities of the survey galaxies is shown 
in Fig.~2. Most of the survey galaxies have velocities of less than 
$v_{\rm{rad}}\leq4000\,\rm{km\,s^{-1}}$. The inclinations of the galaxies 
have been determined using the following formula \citep{Hu26}

\begin{equation}
\cos^2 i = \frac{q^2 - q_0^2}{1 - q_0^2} 
\end{equation}   

\noindent where $q=b/a$ denotes the axial ratio of the galaxy, and $q_0$ is
an intrinsic flattening parameter of the ellipsoid representing the galaxy.
A value of $q_0$=0.20 has been derived by \citet{Ho46}, which is often
adopted. 

\vspace{0.2cm}
\begin{figure}[h]
\begin{center}
\hspace{0.1cm}
\rotatebox{270}{\resizebox{6.0cm}{!}{\includegraphics{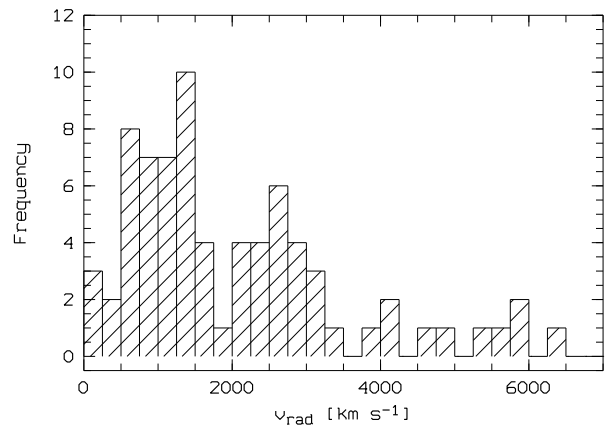}}}
\caption[]{\small Histogram showing the distribution of the radial velocities 
for the 74 survey galaxies.}
\end{center}
\end{figure}
\vspace{0.2cm}    

In Fig.~3 we present a histogram of the distribution of the FIR flux ratio 
($S_{60}/S_{100}$) for the 62 IRAS detected survey galaxies, as well as for 
the LH95 starburst sample, clearly showing the parameter space covered in 
the LH95 study ($S_{60}/S_{100}\geq0.4$), and our survey galaxies peaking at 
values of 0.2--0.35.

\begin{figure}[h]
\begin{center}
\hspace{0.1cm}
\rotatebox{270}{\resizebox{6.0cm}{!}{\includegraphics{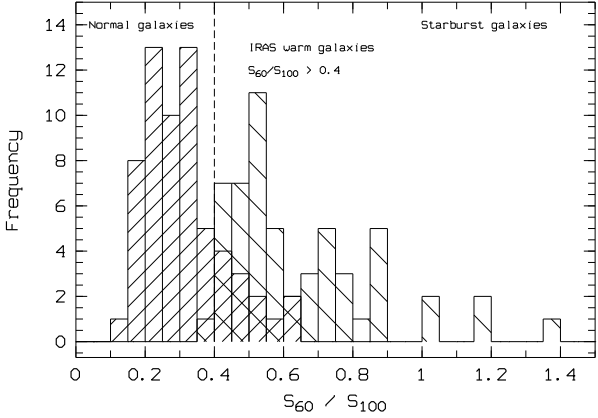}}}
\caption[]{\small Histogram showing the $S_{60}/S_{100}$ distribution of 
the combined samples (starburst sample by \citet{Le95} and our 
non--starburst sample). The vertical dashed line represents the division 
between normal and IRAS warm galaxies.}
\end{center}
\end{figure}

In Fig.~4 we have plotted the distribution of the FIR luminosity of the 
62 IRAS detected survey galaxies, logarithmically in units of solar 
luminosities, ranging from a few $10^{8}\,\rm{L_{\sun}}$ to $10^{11}\,
\rm{L_{\sun}}$. The FIR luminosities have been calculated according to 
\citep{Lo89}

\begin{equation}
L_{\rm{FIR}} = 3.1 \times 10^{39}\,D^2\,[2.58\,S_{60} + S_{100}],
\end{equation}

\noindent where $D$ is the distance to the galaxy in Mpc, $S_{60}$ and 
$S_{100}$ are the FIR flux densities in Jansky at 60$\mu$m and 100$\mu$m, 
respectively.

\begin{figure}[h]
\begin{center}
\rotatebox{270}{\resizebox{6.0cm}{!}{\includegraphics{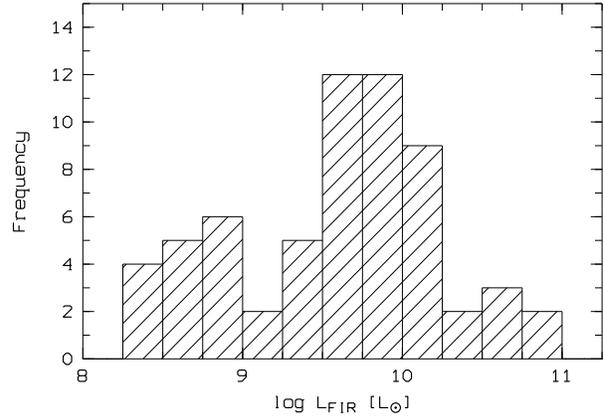}}} 
\caption[]{\small Histogram showing the distribution of the FIR luminosity 
(logarithmically), expressed in units of solar luminosities 
(${\rm L_{\odot}}$) of the 62 galaxies of our survey with IRAS detections.}
\end{center}
\end{figure}


\section{Physical properties of the survey galaxies}

\subsection{Diagnostic DIG diagrams}

In Fig.~5. we have plotted the Diagnostic DIG Diagram \citep[cf.][]{Ro00} 
for our survey galaxies, together with the starburst sample studied by LH95. 
The FIR color ($S_{60}/S_{100}$) for the 62 edge--on galaxies of our 
H$\alpha$ survey, which have been detected with IRAS is plotted as a function 
of the FIR luminosity per unit area. The filled symbols indicate galaxies 
where eDIG has been detected. They separate clearly from the eDIG 
non--detections (denoted by open squares) in this diagram. All IRAS warm 
galaxies ($S_{60}/S_{100} \geq 0.4$) show eDIG. This is valid for the LH95 
sample, as well as for our investigated galaxies, even though they have much 
lower values of $L_{\rm{FIR}}/D^2_{25}$. These targets are mainly Sd type 
spirals. The $S_{60}/S_{100}$ is a measure of the warm dust temperature, and 
is described by interstellar dust models \citep[e.g.,\,][]{Des90}. 

\begin{figure}[h]
\begin{center}
\hspace{0.1cm}
\rotatebox{270}{\resizebox{6.0cm}{!}{\includegraphics{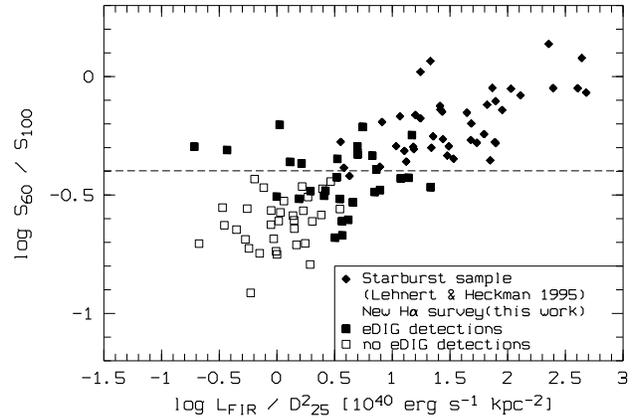}}}
\caption[]{\small Diagnostic DIG diagram for all survey galaxies with IRAS 
detections (62), together with the starburst galaxies, studied by 
\citet{Le95}. Here the ratio of the 60$\mu$m and 100$\mu$m fluxes 
($S_{60}/S_{100}$) is plotted versus the SFR per unit area 
($L_{\rm{FIR}}/D^2_{25}$), expressed in units of 
$10^{40}\rm{\,erg\,s^{-1}\,kpc^{-2}}$. The horizontal dashed line marks the 
threshold for IRAS warm galaxies at $S_{60}/S_{100} \geq 0.4$.} 
\end{center}
\end{figure}

As we zoom into the parameter space of our studied galaxies (see Fig.~6), 
the separation can be seen in more detail. We have also included in this 
plot the positions of galaxies studied by other researchers \citep[e.g.,\,][]
{Pi94,Ra96,Ho99}. These galaxies with/without eDIG detections are denoted by 
filled/open triangles, respectively. Two galaxies (NGC\,973 and NGC\,5403) 
studied by \citet{Pi94}, however, seem to have discordant values in this 
diagram. They claim non--detections for these two objects, but refrain from 
presenting the images. These two galaxies need to be re--observed by more 
sensitive observations to either justify their findings, or to disprove their 
claims, in case there is eDIG present. 

\begin{figure}[h]
\begin{center}
\rotatebox{270}{\resizebox{6.0cm}{!}{\includegraphics{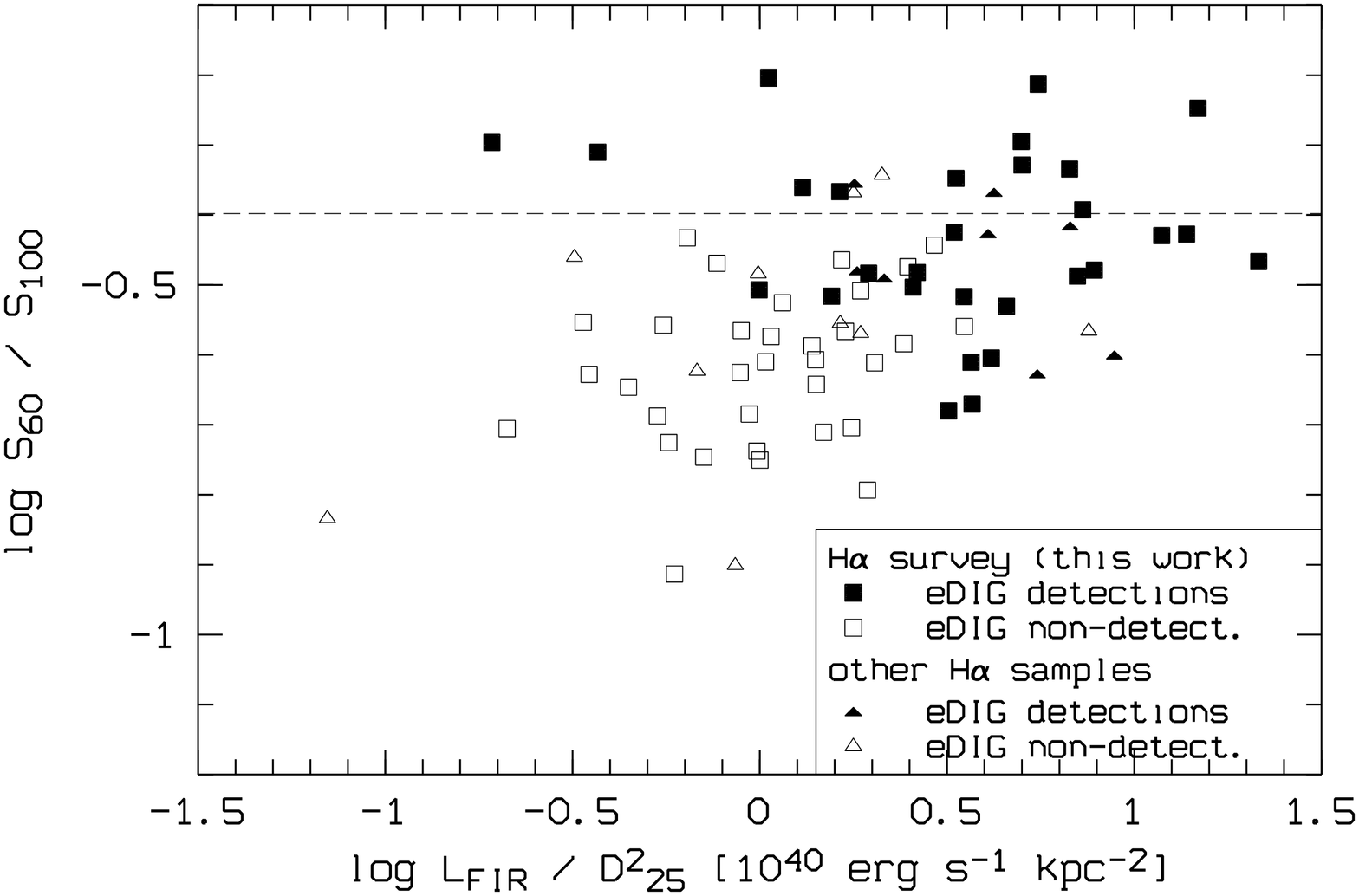}}} 
\caption[]{\small Diagnostic DIG diagram for all survey galaxies with IRAS 
detections (62), shown together with the positions of 18 additional IRAS 
detected galaxies, studied by other investigators 
\citep[i.e.,\,][]{Pi94,Ra92,Ra96,Ho99}. Again, the ratio of the 60$\mu$m and 
100$\mu$m fluxes ($\rm{S_{60}/S_{100}}$) is plotted versus the SFR per unit 
area ($L_{\rm{FIR}}/D^2_{25}$), expressed in units of 
$10^{40}\rm{\,erg\,s^{-1}\,kpc^{-2}}$. The horizontal dashed line marks the 
threshold for IRAS warm galaxies at $\rm{S_{60}/S_{100}} \geq 0.4$.} 
\end{center}
\end{figure}

A minimal SFR per unit area based on the FIR properties can be obtained from 
this diagram, as the detections with the lowest $L_{\rm{FIR}}/D^2_{25}$ ratio 
at low ($S_{60}/S_{100}$) ratios are at $(3.2\pm0.5)\times10^{40}\rm{\,
erg\,s^{-1}\,kpc^{-2}}$. Positive detections at smaller values of 
$L_{\rm{FIR}}/D^2_{25}$ are reached only at considerably higher 
$S_{60}/S_{100}$ ratios of $\sim0.3$, where we define a threshold of 
$1\times10^{40}\rm{\,erg\,s^{-1}\,kpc^{-2}}$. Even {\em fainter} values of 
$L_{\rm{FIR}}/D^2_{25}$ {\em with} eDIG detections are only found among 
{\em IRAS warm galaxies}, such as in the case of ESO\,274--1, which has 
the {\em lowest} $L_{\rm{FIR}}/D^2_{25}$ ratio of our studied galaxies, 
but also {\em one of the highest} $S_{60}/S_{100}$ ratios among our targets 
\citep{Ro01,RoDe01}.


\subsection{Energy input rate per unit area}

In an investigation of a small sample of edge--on galaxies \cite{Da95} found 
evidence for the presence of radio halos for galaxies which exceed a certain 
threshold of energy input per unit area $\rm{\dot{E}_A}$. We follow this 
approach, and calculate $\rm{\dot{E}_A}$ for our survey galaxies, based on 
FIR properties. This energy input rate can be written as 

\begin{equation}
\rm{\dot{E}_A \equiv \frac{dE_{SN}}{dt}/A_{SF} = \nu_{SN}\,E_{SN} / 
\pi\,r^2_{SF}},
\end{equation}

\noindent with $\rm{E_{SN}}$ as the energy released by a supernova (energy 
input per SN), for which we use $\rm{E_{SN}}=10^{51}$\,erg. $\rm{\nu_{SN}}$  
is the SN rate, in units of [$\rm{yr^{-1}}$]. $\rm{A_{SF}}$ denotes the area 
of the SF activity, which is proportional to the SF disk radius squared. We 
use the definition given in \citet{Co92} to derive the SN rate, which is 
$\rm{\nu_{SN} \sim 0.041\,SFR(M > 5\,M_{\odot})}$. The SFRs have been 
calculated according to \citet{Ke98} which can be written in the following 
expression $SFR\,[\rm{M_{\odot}\,yr^{-1}}] = 4.5 \times 10^{-44}\,
L_{\rm{FIR}}\,[\rm{erg\,s^{-1}}]$. The SF disk radius has been measured from 
our H$\alpha$ images. The energy injection rates per unit area are calculated 
in Table~1 (including 4 galaxies of a sub--sample cf.~Paper II) and found to 
range from 1.4$\times10^{-4}$ to 5.9$\times10^{-3}\,{\rm{erg\,s^{-1}\,
cm^{-2}}}$. In Fig.~7 we have plotted the $S_{60}/S_{100}$ ratio as a 
function of the energy injection per unit area ($\dot{E}_A$), coded with 
symbols for detections and non--detections. We have compared our calculated 
SN rates with the ones derived from radio properties by \citet{Da95} for 
e.g., NGC\,891, which are in reasonable agreement. 

\begin{figure}[h]
\begin{center}
\rotatebox{270}{\resizebox{6.0cm}{!}{\includegraphics{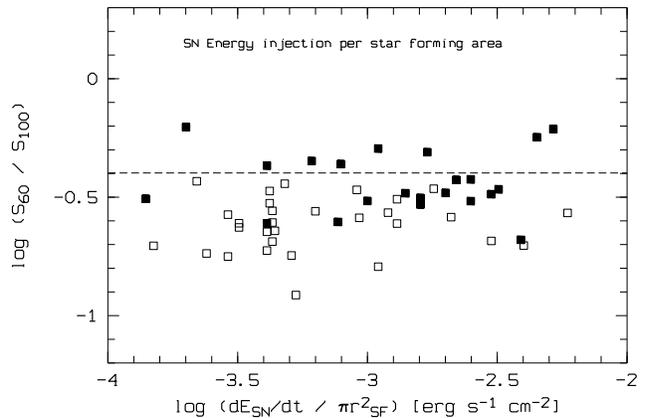}}} 
\caption[]{\small Energy injection rates as a function of the ratio of the 
FIR fluxes at 60$\mu$m and 100$\mu$m ($S_{60}/S_{100}$), both shown 
logarithmically. Data points, representing eDIG detections are coded by 
filled squares and non--detections are denoted by open squares. The 
horizontal dashed line marks, as in previous figures, the threshold for IRAS 
warm galaxies.}
\end{center}
\end{figure}

The diagram above (Fig.~7), however, does not show a clear correlation as in 
the case for the FIR luminosity per unit area. The energy injection as a 
function of $S_{60}/S_{100}$ reveals a larger scatter. There is one 
discordant data point (NGC\,7064), which has a much too low $\dot{E}_A$ 
value for its high $S_{60}/S_{100}$ ratio, given the fact that NGC\,7064 
shows eDIG presence. If that data point is not taken into account, there is 
a slight correlation visible, where galaxies with large values of $\dot{E}_A$ 
also have a large $S_{60}/S_{100}$ ratio, and smaller $\dot{E}_A$ tends to 
have also a smaller $S_{60}/S_{100}$ ratio. In a study of a much smaller 
sample of 16 galaxies \citet{Ir99} came up with a better correlation, 
although one data point did also show a discordant value. However, it should 
be stressed that their parameter space was substantially smaller, and if we 
plot our galaxies covering this particular parameter space, the correlation 
comes out significantly better. The scatter in the energy injection can be 
explained by the fact that galaxies with large intrinsic SF radii tend to 
have low values of $\dot{E}_A$, but strong local SF region can also show 
outflows and thus a mixture of galaxies with low and high values of 
$\dot{E}_A$ is the result. 

\begin{table}[h]
\setcounter{table}{0}
\caption[Star formation rate properties of the survey galaxies]{Star 
formation rate properties of the survey galaxies}
\begin{flushleft}
\begin{center}
\begin{tabular}{lcccc}
\noalign{\smallskip}
\hline\hline
Galaxy & SFR$_{\rm FIR}$ & $\rm{\nu_{SN}}$ & $\rm{r_{SF}}$ & 
$\frac{dE^{\rm{tot}}_{\rm{SN}}}{dt}/A_{\rm{SF}}$ \\ 
& [${\rm M_{\odot}\,yr^{-1}}$] & [${\rm yr^{-1}}$] & [${\rm kpc}$] & 
[${\rm{erg\,s^{-1}\,cm^{-2}}}$] \\ \hline 
NGC\,24 & \,~0.05 & 0.002 & ~\,1.55 & $9.1\times10^{-4}$ \\
NGC\,100 & \,~0.03 & 0.001 & ~\,2.91 & $1.5\times10^{-4}$ \\
UGC\,260 & \,~0.74 & 0.030 & ~\,3.57 & $2.5\times10^{-3}$ \\
ESO\,540--16 & \,~0.08 & 0.003 & ~\,3.94 & $2.2\times10^{-4}$ \\
MCG-2-3-16 & & & ~\,1.34 &  \\
NGC\,360 & \,~0.85 & 0.035 & ~\,5.86 & $1.1\times10^{-3}$ \\
NGC\,669 & \,~1.48 & 0.061 & 14.17 & $3.2\times10^{-4}$ \\
UGC\,1281 & & & ~\,1.67 & \\
NGC\,891 & \,~2.15 & 0.089 & ~\,4.93 & $3.9\times10^{-3}$ \\
UGC\,2082 & \,~0.04 & 0.002 & ~\,2.34 & $3.2\times10^{-4}$ \\
IC\,1862 & \,~5.40 & 0.221 & 23.59 & $4.2\times10^{-4}$ \\
NGC\,1247 & \,~3.56 & 0.146 & 18.05 & $4.8\times10^{-4}$ \\
ESO\,117--19 & \,~2.81 & 0.115 & 13.95 & $6.3\times10^{-4}$ \\
IC\,2058 & \,~0.14 & 0.006 & ~\,3.83 & $4.2\times10^{-4}$ \\
ESO\,362--11 & \,~0.71 & 0.029 & ~\,3.90 & $2.0\times10^{-3}$ \\
ESO\,121--6 & \,~1.07 & 0.044 & ~\,3.94 & $3.0\times10^{-3}$ \\
NGC\,2188 & \,~0.09 & 0.004 & ~\,3.09 & $4.1\times10^{-4}$ \\
ESO\,209--9 & \,~1.49 & 0.061 & ~\,6.30 & $1.6\times10^{-3}$ \\
UGC\,4559 & \,~0.27 & 0.011 & ~\,3.15 & $1.2\times10^{-3}$ \\
NGC\,2654 & \,~0.12 & 0.005 & ~\,3.59 & $4.1\times10^{-4}$ \\
NGC\,2683 & \,~0.16 & 0.007 & ~\,1.32 & $4.0\times10^{-3}$ \\
NGC\,3003 & \,~0.88 & 0.036 & ~\,4.68 & $1.8\times10^{-3}$ \\
NGC\,3221 & 15.39 & 0.631 & 17.64 & $2.2\times10^{-3}$ \\
NGC\,3365 & & & ~\,1.12 & \\
NGC\,3501 & & & ~\,3.60 & \\
NGC\,3600 & \,~0.08 & 0.003 & ~\,2.10 & $7.9\times10^{-4}$\\
NGC\,3628 & ~\,1.88 & 0.077 & &  \\
NGC\,3877 & \,~0.64 & 0.026 & ~\,6.00 & $7.7\times10^{-4}$ \\
NGC\,3936 & \,~0.98 & 0.040 & ~\,5.73 & $1.3\times10^{-3}$ \\
ESO\,379--6 & ~\,1.49 & 0.061 & 12.62 & $4.1\times10^{-4}$ \\
NGC\,4206 & \,~0.16 & 0.007 & ~\,4.01 & $4.3\times10^{-4}$ \\
NGC\,4216 & \,~0.40 & 0.016 & ~\,5.74 & $5.3\times10^{-4}$ \\
NGC\,4235 & \,~0.21 & 0.009 & ~\,2.29 & $1.7\times10^{-3}$ \\
NGC\,4256 & \,~0.51 & 0.021 & ~\,6.59 & $5.1\times10^{-4}$\\
NGC\,4388 & \,~7.20 & 0.295 & ~\,7.75 & $5.2\times10^{-3}$ \\
NGC\,4700 & \,~1.00 & 0.041 & ~\,6.25 & $1.1\times10^{-3}$ \\
NGC\,4945 & 13.10 & 0.537 & \hspace*{-0.39cm} $\geq$11.21 & \hspace*{-0.50cm} 
$\geq4.5\times10^{-3}$\\
NGC\,5290 & \,~1.58 & 0.065 & ~\,6.50 & $1.6\times10^{-3}$ \\
NGC\,5297 & \,~8.60 & 0.353 & ~\,8.00 & $5.9\times10^{-3}$\\
NGC\,5775 & \,~8.42 & 0.345 & 10.72 & $3.2\times10^{-3}$ \\
ESO\,274--1 & \,~0.04 & 0.002 & &  \\
NGC\,5965 & \,~0.96 & 0.039 & ~\,9.89 & $4.3\times10^{-4}$ \\
NGC\,6722 & & & 21.43 & \\
IC\,4837A & \,~2.24 & 0.092 & ~\,6.79 & $2.1\times10^{-3}$ \\
ESO\,142--19 & & & &  \\
IC\,4872 & & & ~\,1.67 &  \\
NGC\,6875A & ~\,0.83 & 0.034 & ~\,6.00 & $1.0\times10^{-3}$ \\
\hline
\end{tabular}
\end{center}
\end{flushleft}    
\end{table} 

\begin{table}[h]
\setcounter{table}{0}
\caption[continued]{continued}
\begin{flushleft}
\begin{center}
\begin{tabular}{lcccc}
\noalign{\smallskip}
\hline\hline
Galaxy & SFR$_{\rm FIR}$ & $\rm{\nu_{SN}}$ & $\rm{r_{SF}}$ & 
$\frac{dE^{\rm{tot}}_{\rm{SN}}}{dt}/A_{\rm{SF}}$ \\ 
& [${\rm M_{\odot}\,yr^{-1}}$] & [${\rm yr^{-1}}$] & [${\rm kpc}$] & 
$[{\rm{erg\,s^{-1}\,cm^{-2}}}$] \\ \hline 
FGC\,2290$^a$ & ~\,2.03 & 0.083 & ~\,5.40 & $3.0\times10^{-3}$ \\
IC\,5052 & \,~0.08 & 0.003 & ~\,4.95 & $1.4\times10^{-4}$ \\
IC\,5071 & \,~1.07 & 0.044 & 10.43 & $4.3\times10^{-4}$ \\
IC\,5096 & \,~0.97 & 0.040 & ~\,9.81 & $4.4\times10^{-4}$ \\
NGC\,7064 & \,~0.06 & 0.002 & ~\,3.60 & $2.0\times10^{-4}$ \\
NGC\,7090 & \,~0.58 & 0.024 & ~\,4.20 & $1.4\times10^{-3}$ \\
UGC\,11841 & & & &  \\
NGC\,7184 & \,~2.22 & 0.091 & 10.22 & $9.3\times10^{-4}$ \\
IC\,5171 & \,~0.75 & 0.031 & ~\,5.10 & $1.3\times10^{-3}$ \\
IC\,5176 & \,~1.47 & 0.060 & ~\,5.11 & $2.5\times10^{-3}$\\
NGC\,7339 & & & ~\,1.16 &  \\
NGC\,7361 & \,~0.15 & 0.086 & ~\,4.75 & $2.9\times10^{-4}$ \\
NGC\,7412A & & & ~\,2.39 &  \\
UGC\,12281 & \,~0.54 & 0.022 & 10.00 & $2.3\times10^{-4}$ \\
NGC\,7462 & \,~0.41 & 0.017 & ~\,5.41 & $6.1\times10^{-4}$ \\
UGC\,12423 & & & &  \\ 
NGC\,7640 & \,~0.12 & 0.005 & ~\,3.56 & $4.1\times10^{-4}$ \\
ESO\,240--11 & \,~1.51 & 0.062 & 14.97 & $2.9\times10^{-4}$ \\
\hline
\end{tabular}
\end{center}
\vspace{0.3cm}
$^a$MCG-01-53-012
\end{flushleft}    
\end{table}


\section{Discussion}

The presence of galactic gaseous halos is usually regarded as the result of 
strong SF activity in the underlying galaxy disks. Although detailed 
kinematic information has mostly eluded us to confirm an associated 
outflow, many pieces of evidence, that have been gathered in the last couple 
of years, strengthen the concept of a disk--halo interaction. 

Alternatively, the gas could also be arising from infalling high--velocity 
clouds (HVCs) in the halo, which are seen in our Milky Way 
\citep[e.g.,\,][]{Wak97} or could maybe originating from galaxy interactions 
or past mergers.
 
From the theoretical perspective several scenarios have been suggested for 
the gas transport from the galactic disk into the halo, among them the {\em 
galactic fountains} \citep{Sh76,Av00}, {\em chimneys} \citep{No89}, {\em 
superbubble outbreak} \citep{Ma99}, or {\em superwinds} \citep{He90}.  

Past observations have indicated that several edge--on galaxies show DIG 
at large extraplanar distances, whereas on the other side examples were 
found where no eDIG has been detected \citep{Ra96}. The present H$\alpha$ 
survey has contributed significantly to the question on the occurrence of 
gaseous halos in normal spiral galaxies. This first systematic approach 
has investigated galaxies with a broad distribution of FIR luminosities. 


\subsection{Galactic gaseous halos -- constraints and dependency}

It is the strength of the SF activity, both locally and on global scales, 
and the high values of the FIR flux ratio S$_{60}$/S$_{100}$ (a measure of 
the warm dust temperature), that determines the presence of extraplanar DIG 
in halos of several edge--on spiral galaxies.

Despite possible differences in the ionization mechanism of the halo gas in 
normal and starburst galaxies (photoionization by hot O and B stars and SNe 
driven superwinds, respectively), it is interesting to note the generally 
good correlation between the SFR per unit area and the FIR flux ratio at 
60\,$\mu$m and 100\,$\mu$m.

The normal galaxies represent the fainter extension of the starburst regime 
with an obvious soft transition. While the SFR per unit area is a good tracer 
for eDIG in this diagram (see Fig.~5), the diagnostic diagram with 
$\dot{E}_A$ shows a less obvious correlation (cf. Sect.~4.2). However, 
individual galaxies which have not been investigated yet in the disk--halo 
interaction (DHI) context can be selected according to their position within 
the DDD, to be used for a comparison study in multifrequency investigations.  

Comparing the observed morphology with data presented by other researchers 
\citep[e.g.,\,][]{Ra96,Ho99}, the results are generally in very good 
agreement. Furthermore most galaxies, including all previous studied 
samples, follow the above mentioned correlation very well. However, it should 
be noted that a very few individual cases show subtle differences 
(cf. Sect.~4.1).

Galaxies with extraplanar DIG either fall in the upper right or in the upper 
left corner region of the diagnostic DIG diagram (DDD). Galaxies with 
detectable, but less prominent gaseous halos populate the region adjacent to 
these regions (middle part of the DDD), whereas galaxies with no detectable 
gaseous halos are located in the lower left corner. In addition to that, 
there are two regions that are not occupied by any galaxies. These are the 
extreme upper left and lower right corners in the DDD. Obviously in those 
parts there is no physical meaningful parameter space for galaxies. In spite 
of the current calculation of the SFR from the FIR, the FIR may be low in 
some cases with still a significant SFR if the amount of dust is small. On 
the other side, a high dust temperature also demands a high star formation 
rate, although high dust temperatures {\em do not} necessarily imply high 
SFRs per unit area. For the subluminous IRAS bright (warm) galaxies strong 
local SF activity with a high $S_{60}/S_{100}$ ratio also leads to strong 
{\em local} gaseous outflows (e.g.,\,ESO\,274--01)! This area is dominated 
by the extreme late--type and irregular galaxies such as ESO\,274--01. 
However, this region is also occupied at least by one observed {\em Seyfert} 
galaxy (NGC\,4235), whose activity (powered by an AGN) is restricted to the 
nuclear region.

In Table~2 all galaxies of this current H$\alpha$ survey are listed in 
order of {\em decreasing} SF rate per unit area ($L_{\rm{FIR}}/D^2_{25}$). 
Those galaxies of our H$\alpha$ survey, which have also been investigated by 
other protagonists, are indicated by an asterisk. We list in Table~2 for 
all galaxies the $S_{60}/S_{100}$ ratio, in addition to the distances of the 
galaxies, the major axis diameter $a_{25}$ in arcminutes, the above 
mentioned SF rate per unit area, and also the observed DIG morphology. As 
previously mentioned, from the total of 74 galaxies there were 12 galaxies 
with no measured FIR fluxes. We have therefore listed them (in order of 
increasing R.A.) at the end of this table.

The galaxy with the {\em highest} SFR per unit area in our sample is 
NGC\,5775, reaching a value of about 21.5 (in units of $\rm{10^{40}\,erg\,
s^{-1}\,kpc^{-2}}$), which was also included in the sample by LH95. We have 
included four galaxies of their sample also into our survey. We have 
calculated the $S_{60}/S_{100}$ ratio and the $L_{\rm{FIR}}/D^2_{25}$ ratio 
for the galaxies of the LH95 sample. The resulting numbers are listed in 
Table~4, where we have marked the five galaxies, that we have studied as 
well, with an asterisk. One further galaxy of the LH95 sample recently has 
been studied by \cite{Co00}, and this galaxy is marked in Table~4 with a 
dagger. Finally, all other normal galaxies studied in the DIG context by 
various researchers have been listed in Table~3, for which we have also 
calculated the diagnostic ratios.

The galaxy with the {\em lowest} SFR per unit area (0.2 in same units) 
in our survey is ESO\,274--01. It shows extended H$\alpha$ emission locally. 
Although this galaxy fails to represent a continuous halo, there are a few 
bright local features, whereas on other positions in the disk there is 
almost no detectable H$\alpha$ emission at all. However, the fact that strong 
local extended emission in this galaxy is detected, is not too surprising, 
as it has one of the {\em highest} $S_{60}/S_{100}$ ratios among our survey 
galaxies. In Table~2 we further indicate whether or not {\em radio thick 
disks} (or radio halos), and X--ray halos have been detected by other 
investigators for the studied galaxies. Positive detections are marked by a 
bullet ($\bullet$), and non--detections are marked by a circle ($\circ$). 
For galaxies where neither one of the two symbols are indicated, no radio 
continuum and/or X--ray observations have been performed yet in search of 
extended emission. 

In Tables~2+3 we further indicate whether or not radio thick disks and X--ray 
halos have been detected for the studied galaxies. If we compare the presence 
of gaseous halos with further evidence of the disk--halo interaction, coming 
from these other wavelength regimes, we generally find a good correlation of 
both the presence of {\em radio thick disks} and extended H$\alpha$ emission 
in several cases. However, it should be stressed that there are still many 
galaxies of our survey, which have not been observed in the radio regime with 
sufficiently high resolution, which would allow a detailed comparison. In the 
X--ray regime there is also the demand for more sensitive observations of 
nearby edge--on galaxies. Besides the detection of X--ray halos in several 
nearby {\em starburst} galaxies, information on X--ray halos in 
non--starburst galaxies are yet still very scarce. NGC\,891 is one of the 
best studied cases in this respect \citep{Br94}. But with the currently 
available more sensitive X--ray telescopes such as XMM/Newton and Chandra 
things will hopefully change soon. A very recent XMM/Newton EPIC pn detection 
of NGC\,3044 confirms the presence of extended soft X--ray emission (Rossa 
et al., in prep.).  

\subsection{Clustered SNe as a cause of galactic halos}

In the theories of chimneys and supperbubble outbreak \citep{No89,Ma88} the 
presence of gaseous halos have their origin in gas ejections from the disk 
into the halo, triggered by clustered and correlated SNe. Whether all 
bubbles, created by clustered SNe, lead to an energy and momentum flux 
sufficiently high enough to entrain gaseous matter well above the 
disk--halo interface (i.e. a blowout) has been investigated by \citet{Hei90}.
\citet{Koo92} have shown that some of the detected {\em worms} of earlier 
works by \cite{Hei84} appear to be ionized. The general picture is that 
OB associations produce supershells in the surrounding ISM and if these 
supershells expand to large scale--heights they appear as worms. Large 
OB associations can produce up to 7000 supernova progenitors \citep{McK97}. 

Following \citet{Hei90} we calculate the number of clustered SNe at the nadir 
of a filament which corresponds to our determined threshold of SF activity 
per unit area. Taking the adjusted relation of \citet{Hei90}

\begin{equation}
N = 222\,L(\rm{obs})_{38}\,\tau_{\tiny{\hbox{H\,\sc{ii}},7}},
\end{equation}

\noindent where $\tau_{\tiny{\hbox{H\,\sc{ii}},7}}$ is the lifetime of a 
cluster's \hbox{H\,\sc{ii}} region in units of $10^{7}$\,yrs, and 
$L(\rm{obs})_{38}$ is the H$\alpha$ luminosity in units of $10^{38}\,
\rm{erg\,s^{-1}}$, we compare this with our derived SF threshold per unit 
area $\dot{E}^{\rm thres}_{A\,25} \equiv L_{\rm{FIR}}/D^2_{25} \leq 
(3.2\pm0.5)\times10^{40}\,\rm{erg\,s^{-1}\,kpc^{-2}}$. This will be only a 
simplistic estimate, as we know that the exact value is difficult to assess 
from many uncertainties and assumptions being made. The expression derived 
by \citet{Hei90} rests on observational data of a sample of Sb galaxies by 
\citet{Ken89}, galaxies which are similar to our studied edge--on spirals. 

We use the conversion factor of the H$\alpha$ to FIR luminosity of 1/176 
given by \citet{Ke98}. Even values of 1/550 have been derived from the 
literature for several late--type spirals. This can be calculated from 
H$\alpha$ data published in the literature \citep[e.g.,\,][]{Ro00,Bos02}, and 
from the determined FIR luminosities \citep[cf.][]{Lo89}. However, in 
those cases (H$\alpha$ measurements of edge--on spirals) there is always an 
uncertainty involved due to extinction by dust. 

With a mean effective area of the SF filament (approximated by 
$A_{\rm fil}=\pi\,r_{\rm fil}^2$), this yields with 
$\tau_{\tiny{\hbox{H\,\sc{ii}},7}}=2$ \citep[cf.][]{Hei90} the following 
expression

\begin{equation}
N = 254\,\dot{E}^{\rm thres}_{\rm 40}\,{r^2_{\rm fil}}\,[\rm{kpc}].
\end{equation}  

\noindent Using our derived minimal SF area in units of $10^{40}\,\rm{erg\,
s^{-1}\,kpc^{-2}}$, multiplied with the area of a filament at the nadir in 
the disk $<{A_{\rm fil}}>$, for IC\,5052, which is $<r_{\rm fil}>=37.5$\,pc, 
this yields $N\approx4$ clustered SNe. 

Although this is a very rough estimation, this number of clustered SNe would 
correspond to our derived SF activity per unit threshold. 

In an alternative simple approach we calculate the number of SNe which 
are corresponding to our derived SF threshold at the nadir of a filament. 
Taking the energy to be released by one supernova and ejected into the 
ambient ISM ($10^{51}$\,erg), we calculate the required number of SNe in a 
time interval of $\tau=10^7$\,yr and per kpc$^2$ following 

\begin{equation}
N/A_{\rm SF}\,[\rm kpc^2] = \frac{\dot{E}^{\rm thres}_{\rm 40}\,\tau}{E_{\rm 
kin}}.
\end{equation}  

Multiplying this with the area of the smallest filament $(A_{\rm fil} = \pi 
r^2_{\rm fil})$ and inserting the determined value of the filament radius 
($r_{\rm fil}$ = 37.5\,pc), this yields $N\approx45$\,SNe.

Concluding, the derived numbers are in reasonable agreement to support the 
working hypothesis that warm gaseous halos are a direct consequence of 
SF driven outflows. Even given the great uncertainty in calculating these 
numbers, it is evident that only a modest number of clustered SNe are 
required in order to yield outflows at the derived threshold. 

The observed morphology of outflows is different from the starburst galaxies, 
at least from nucleated starbursts, as in almost no cases a nuclear 
outflowing cone is detected. The outflowing structures (i.e. filaments) 
emanate from strong SF regions in the disk, which are distributed along the 
disk (not necessarily confined to the nuclear regions). It has to be 
emphasized, however, that there may also be a small contribution of infalling 
gas (e.g., from HVCs). Unless kinematic information is available, which has 
up to now mostly eluded us, it is hard to predict the fraction of the latter 
contribution.
							 

\section{Summary}

We have introduced the selection criteria for our H$\alpha$ survey, 
consisting of 74 nearby ($D\leq85$\,Mpc), mostly late--type, edge--on spiral 
galaxies with inclinations of $i\geq76^\circ$. It is the largest H$\alpha$ 
survey to date, investigating extraplanar DIG. We have shown some optical and 
FIR properties of the selected galaxies. 

Our survey galaxies populate the fainter end of the FIR luminosity 
distribution, in contrast, to starburst galaxies which populate the high 
luminosity range. The parameter space of the $S_{60}/S_{100}$ ratio for our 
galaxies is also filled in the lower range, as most galaxies fall in between 
values of 0.2 and 0.4. We covered also an overlap area of the starburst 
galaxies. The starburst sample by LH95 has values of $\geq0.4$ (IRAS warm 
galaxies). The FIR luminosities of our studied galaxies range from a few 
$10^{8}$\,$\rm{L_{\odot}}$ up to $10^{11}$\,$\rm{L_{\odot}}$. Whereas the 
SFR per unit area in starburst galaxies have considerably higher values of 
up to 480 in units of $10^{40}\rm{\,erg\,s^{-1}\,kpc^{-2}}$, the values for 
our galaxies range from 0.2--21.5 in the same units. We have detected eDIG 
in 30 of our studied galaxies. It can be concluded that, despite the fact 
that eDIG is {\em not} encountered {\em in all} late--type spirals, however, 
a significant fraction of our studied galaxies (41\,\%) possesses eDIG. 
Therefore, eDIG seems to be omnipresent in late--type spirals with high SF 
activity above $(3.2\pm0.5)\times10^{40}\rm{\,erg\,s^{-1}\,kpc^{-2}}$. There 
is, however, also a dependence on the $S_{60}/S_{100}$ ratio, and therefore 
the minimal threshold is also a function of $S_{60}/S_{100}$, as already 
evidenced in our study of a sub--sample \citep{Ro00}.  

In our second paper \citep{RoDe01} the actual results for the individual 
galaxies are presented. There we will present the detailed morphology of 
eDIG of the studied galaxies, and in particular show several cases of 
galaxies with extended warm gaseous halos. Concluding, our working hypothesis 
is further justified and corroborated, that the presence of gaseous halos in 
late--type spirals is a direct consequence of SF activity in the underlying 
galactic disk. 


\begin{table*}[h]
\setcounter{table}{1}
\centering
\begin{minipage}{18cm}\small
\caption[]{Important physical parameters and DIG properties of the H$\alpha$ 
survey galaxies}
\begin{tabular}{lcccccccc}
\noalign{\smallskip}
\hline\hline
Galaxy & $S_{60}/S_{100}$ & $D$ & $a_{25}$ & $L_{\rm{FIR}}/D^2_{25}$ & 
DIG morph.$^{\footnotesize a}$ & Radio halo & X--ray halo & Refs.\\ 
& & [Mpc] & [$'$] & [$\rm{10^{40}\,erg\,s^{-1}\,kpc^{-2}}$] & & (thick disk) 
& & \\
\hline
NGC\,5775$^\star$ & 0.3415 & 26.7 & \,~3.8 & {\bf 21.49} & h$_{\rm b}$,f,pl 
& $\bullet$ & $\bullet$ & (6); (11); (12)\\
NGC\,4945$^\star$ & 0.5661 & \,~7.5 & 20.4 & {\bf 14.79} & f,h$_{\rm b}$,pl 
& $\bullet$ & & \\
NGC\,3221 & 0.3738 & 54.7 & \,~3.1 & {\bf 13.80} & ee,f,h$_{\rm f}$ & 
$\bullet$ & & (10)\\
NGC\,4634 & 0.3720 & 19.1 & \,~2.6 & {\bf 11.84} & ee,f,h$_{\rm b}$,pa & 
& & \\
NGC\,4402 & 0.3320 & 22.0 & \,~3.9 & {\bf \,~7.82} & ee,f,pa & & & \\
IC\,2135$^b$ & 0.4051 & 17.7 & \,~2.8 & {\bf \,~7.28} & e,h,p & & & \\
ESO\,121--6 & 0.3256 & 16.2 & \,~3.9 & {\bf \,~7.05} & h$_{\rm f}$ & & & \\
NGC\,3044$^\star$ & 0.4633 & 17.2 & \,~4.9 & {\bf \,~6.72} & ee,f,h$_{\rm b}$ & $\bullet$ & $\bullet$ & (8);(13) \\
NGC\,4388$^\star$ & 0.6127 & 33.6 & \,~5.5 & {\bf \,~5.53} & f,h$_{\rm f}$,pa & $\bullet$ & & \\
NGC\,3628$^\star$ & 0.4694 & \,~7.7 & 12.9 & {\bf \,~5.02} & ee,f,pl & 
$\bullet$ & $\bullet$ & (4)\\
NGC\,4700 & 0.5074 & 24.0 & \,~3.0 & {\bf \,~4.99} & h$_{\rm b}$,f,pa & 
$\bullet$ & & (5)\\
ESO\,209--9 & 0.2948 & 14.9 & \,~6.2 & {\bf \,~4.56} & h$_{\rm b}$,pa,pl & 
& & \\
NGC\,3877 & 0.2486 & 12.1 & \,~5.3 & {\bf \,~4.15} & ee,f,pl & & & \\
NGC\,4302$^\star$ & 0.2136 & 18.8 & \,~5.5 & {\bf \,~3.69} & ee,pl & 
$\bullet$ & & (9)\\
ESO\,379--6 & 0.2452 & 39.8 & \,~2.6 & {\bf \,~3.67} & ee,pl & & & \\
ESO\,117--19 & 0.2759 & 71.4 & \,~2.0 & {\bf \,~3.51} & d & & & \\
IC\,5176 & 0.3044 & 23.3 & \,~4.5 & {\bf \,~3.51} & ee,h$_{\rm f}$ & 
$\bullet$ & & (7)\\
NGC\,7462 & 0.4492 & 14.1 & \,~4.0 & {\bf \,~3.34} & f,h$_{\rm f}$,pl & 
$\bullet$ & & (5)\\
UGC\,260 & 0.3761 & 28.5 & \,~2.7 & {\bf \,~3.30} & eh2,f,h$_{\rm f}$,pec & 
& & \\
NGC\,891$^\star$ & 0.2089 & \,~9.5 & 14.0 & {\bf \,~3.19} & ee,eh2,f,h$_{\rm 
b}$ & $\bullet$ & $\bullet$ & (1); (3)\\
NGC\,1247 & 0.3604 & 52.6 & \,~3.4 & {\bf \,~2.92} & d,pa & & & \\
ESO\,362--11 & 0.3295 & 17.9 & \,~4.7 & {\bf \,~2.63} & h$_{\rm f}$ & & & \\
NGC\,5290 & 0.3139 & 34.4 & \,~3.7 & {\bf \,~2.57} & ee,f,h$_{\rm b}$ & & & \\
IC\,1862 & 0.3357 & 85.3 & \,~2.8 & {\bf \,~2.48} & d,n & & & \\
IC\,4837A & 0.2605 & 37.9 & \,~4.1 & {\bf \,~2.42} & d & & & \\
NGC\,3936 & 0.2447 & 29.0 & \,~3.9 & {\bf \,~2.03} & d,n & & & \\
NGC\,7090 & 0.3287 & 11.4 & \,~7.7 & {\bf \,~1.95} & a,ee,f,h$_{\rm f}$ & 
$\bullet$ & & (5)\\
NGC\,360 & 0.1609 & 30.7 & \,~3.5 & {\bf \,~1.94} & d,ee,pl & & & \\
IC\,5171 & 0.3101 & 34.1 & \,~3.0 & {\bf \,~1.86} & d & & & \\
NGC\,2683 & 0.1976 & \,~5.5 & \,~8.8 & {\bf \,~1.76} & d,n & $\bullet$ & 
& (10)\\
NGC\,5297 & 0.2714 & 67.5 & \,~5.4 & {\bf \,~1.69} & d,n & $\bullet$ & 
& (10)\\
NGC\,3003$^\star$ & 0.3434 & 19.7 & \,~6.0 & {\bf \,~1.65} & d,n & & & \\
NGC\,2188$^\star$ & 0.4301 & \,~7.9 & \,~4.7 & {\bf \,~1.63} & f,h$_{\rm f}$,pl& & & \\
NGC\,6875A & 0.3048 & 42.4 & \,~2.8 & {\bf \,~1.56} & ee?,pa & & & \\
IC\,4351 & 0.1947 & 35.5 & \,~6.0 & {\bf \,~1.48} & n & & & \\
IC\,5096 & 0.2280 & 41.9 & \,~3.2 & {\bf \,~1.42} & n & & & \\
IC\,5071 & 0.2472 & 41.6 & \,~3.4 & {\bf \,~1.41} & d & & & \\
NGC\,7184 & 0.2588 & 34.1 & \,~6.0 & {\bf \,~1.38} & d,n & & & \\
NGC\,3600 & 0.4362 & \,~9.6 & \,~4.1 & {\bf \,~1.30} & ee & $\circ$ & & (9) \\
IC\,2058 & 0.2982 & 18.2 & \,~3.1 & {\bf \,~1.15} & n & & & \\
NGC\,7361 & 0.2669 & 15.9 & \,~3.8 & {\bf \,~1.07} & d,n & & & \\
NGC\,7064 & 0.6250 & 11.4 & \,~3.4 & {\bf \,~1.06} & ee,h$_{\rm f}$,pl & 
$\bullet$ & & (7)\\
NGC\,669 & 0.2455 & 62.4 & \,~3.1 & {\bf \,~1.04} & n & & & \\
ESO\,240--11 & 0.1776 & 37.9 & \,~5.3 & {\bf \,~1.00} & n & $\bullet$ & 
& (7)\\
IC\,5052 & 0.3112 & \,~7.9 & \,~5.9 & {\bf \,~1.00} & a,ee,h$_{\rm b}$,pl & 
& & \\
UGC\,12281$^\star$ & 0.1830 & 34.2 & \,~3.5 & {\bf \,~0.98} & n & & & \\
MCG-01-53-012 & 0.2067 & 79.5 & \,~3.0 & {\bf \,~0.94} & n & & & \\
UGC\,4559 & 0.2721 & 27.8 & \,~3.2 & {\bf \,~0.89} & n & & & \\
UGC\,10288$^\star$ & 0.2370 & 27.3 & \,~4.8 & {\bf \,~0.89} & n & $\bullet$ 
& & (9)\\
NGC\,24 & 0.3396 & \,~6.8 & \,~5.8 & {\bf \,~0.77} & d,n & & & \\
NGC\,4256 & 0.1794 & 33.7 & \,~4.1 & 
{\bf \,~0.71} & d,n & & & \\
ESO\,540--16 & 0.3692 & 20.7 & \,~2.7 & 
{\bf \,~0.64} & d & & & \\
NGC\,4216 & 0.1221 & 16.8 & \,~7.9 & 
{\bf \,~0.59} & d,n & & & \\
NGC\,2654 & 0.1882 & 17.9 & \,~4.2 & 
{\bf \,~0.57} & d,n & & & \\
NGC\,5965 & 0.2769 & 45.5 & \,~4.7 & 
{\bf \,~0.55} & d,n & & & \\
NGC\,4206 & 0.2054 & 16.8 & \,~5.2 & 
{\bf \,~0.53} & d,pa & & & \\
NGC\,7640 & 0.2259 & \,~8.6 & \,~9.9 & 
{\bf \,~0.45} & d,n & & & \\
NGC\,4235 & 0.4896 & 32.1 & \,~3.8 & 
{\bf \,~0.37} & ee,h$_{\rm f}$ & $\bullet$ & & (2)\\
\hline
\end{tabular}
\end{minipage}
\end{table*}

\clearpage

\begin{table*}[h]
\setcounter{table}{1}
\centering
\begin{minipage}{20cm}\small
\caption[continued]{continued}
\begin{tabular}{lcccccccc}
\noalign{\smallskip}
\hline\hline
Galaxy & $S_{60}/S_{100}$ & $D$ & $a_{25}$ & $L_{\rm{FIR}}/D^2_{25}$ & 
DIG morph.$^{\footnotesize a}$ & Radio halo & X--ray halo & Refs.\\ 
& & [Mpc] & [$'$] & [$\rm{10^{40}\,erg\,s^{-1}\,kpc^{-2}}$] & & (thick disk) 
& & \\
\hline
UGC\,2082 & 0.2356 & \,~9.4 & \,~5.4 & 
{\bf \,~0.35} & d,n & & & \\
NGC\,5170 & 0.2796 & 20.0 & \,~8.3 & 
{\bf \,~0.34} & n & & & \\
NGC\,100 & 0.1971 & 11.2 & \,~5.6 & 
{\bf \,~0.21} & n & & & \\
ESO\,274--1 & 0.5055 & \,~7.0 & 11.0 & 
{\bf \,~0.19} & ee,f & & & \\
\hline
MCG-2-3-16 & -- & 17.9 & \,~3.0 & 
{\bf \,~} & a,n & & & \\
UGC\,1281 & -- & \,~4.6 & \,~4.5 & 
{\bf \,~} & n & $\circ$ & & (9)\\
IC\,2531 & -- & 33.0 & \,~6.9 & 
{\bf \,~} & pl & & & \\
NGC\,3365 & -- & 13.2 & \,~4.4 & 
{\bf \,~} & d,n & & & \\
NGC\,3501 & -- & 15.1 & \,~3.5 & 
{\bf \,~} & n & $\circ$ & & (9)\\
NGC\,6722 & -- & 76.7 & \,~2.9 & 
{\bf \,~} & n & & & \\
ESO\,142--19 & -- & 56.2 & \,~4.4 & 
{\bf \,~} & n & & & \\
IC\,4872 & -- & 25.7 & \,~3.5 & 
{\bf \,~} & n & & & \\
UGC\,11841 & -- & 79.9 & \,~2.8 & 
{\bf \,~} & n & $\circ$ & & (9)\\
NGC\,7339 & -- & 17.9 & \,~2.7 & 
{\bf \,~} & n& & & \\
NGC\,7412A & -- & 12.4 & \,~3.5 & 
{\bf \,~} & a,n & & & \\
UGC\,12423 & -- & 64.5 & \,~3.4 & 
{\bf \,~} & n & $\circ$ & & (9)\\
 &  &  &  &  & & & & \\
\hline
\end{tabular}
\end{minipage}
\hspace*{-1.5cm}{\footnotesize $^{a}$ a$=$arc(s), d$=$disk emission (only 
planar DIG), ee$=$extended emission (locally), eh2$=$extraplanar 
\hbox{H\,{\sc ii}} region(s),\\ 
\hspace*{-2.3cm} f$=$filament(s), h$_{\rm b}$=bright halo, h$_{\rm f}$=faint 
halo, n$=$no (e)DIG, pa$=$patch(es), pec$=$peculiar, pl$=$plume(s), \\}
\vspace*{0.2cm}
\hspace*{-0.2cm}{\footnotesize References: (1) Bregman \& Pildis (1994); (2) 
Colbert et al.~(1996); (3) Dahlem et al.~(1994); (4) Dahlem et al.~(1996);\\ 
\hspace*{-0.28cm} (5) Dahlem et al.~(2001); (6) Duric et al.~(1998); (7) 
Harnett \& Reynolds (1991); (8) Hummel \& van der Hulst (1989);\\ 
\hspace*{-2.3cm} (9) Hummel et al.~(1991); (10) Irwin et al.~(1999); (11) 
T\"ullmann et al.~(2000); (12) Rossa et al.~(2002);\\ 
\hspace*{-2.1cm} (13) Rossa et al., in prep., $^{b}$IC\,2135 was formerly 
known as NGC\,1963 ($\bullet$ = detection, $\circ$ = non--detection)} 
\end{table*}

\begin{table*}[h]
\caption[]{Galaxies studied by other investigators}
\begin{flushleft}
\begin{tabular}{lcccccccl}
\noalign{\smallskip}
\hline\hline
Galaxy & $S_{60}/S_{100}$ & $D$ & $a_{25}$ &$L_{\rm{FIR}}/D^2_{25}$ & DIG 
morphology$^{\footnotesize a}$ & Radio halo & X--ray halo & Refs.\\ 
 & & [Mpc] & [$'$] & [$\rm{10^{40}\,erg\,s^{-1}\,kpc^{-2}}$] & & (thick 
disk) & & \\
\hline
NGC\,4217 & 0.2511 & 17.0 & ~\,4.9 & 8.85 & pl,pa & & & (9) \\
NGC\,5403 & 0.2717 & 37.6 & ~\,3.0 & 7.56 & n & & & (7) \\
NGC\,3556 & 0.3841 & 14.1 & ~\,8.1 & 6.74 & ee,h$_{\rm f}$? & $\bullet$ & 
$\bullet$ & (1);(10);(11)\\
NGC\,4013 & 0.2360 & 17.0 & ~\,4.8 & 5.51 & ee,f,h$_{\rm f}$ & & & (9) \\
NGC\,4631 & 0.4292 & ~\,6.9 & 14.8 & 4.22 & f,h$_{\rm f}$,pa & $\bullet$ & 
$\bullet$ & (5); (8) \\
NGC\,5777 & 0.3745 & 30.5 & ~\,3.1 & 4.07 & pl & & & (7) \\
NGC\,4522 & 0.3235 & 16.8 & ~\,3.5 & 2.15 & ee,eh2,f & & & (6) \\
NGC\,973 & 0.4541 & 66.1 & ~\,3.7 & 2.12 & n & & & (7) \\
UGC\,3326 & 0.2696 & 56.4 & ~\,3.4 & 1.86 & pl & & & (7) \\
NGC\,3432 & 0.3313 & ~\,7.8 & ~\,6.9 & 1.82 & ee,eh2 & & & (2) \\
NGC\,55 & 0.4423 & ~\,1.3 & 27.5 & 1.79 & ee,eh2,f,pl,& & $\circ$ & (3);(4);
(14)\\
NGC\,4762 & 0.4286 & 16.8 & ~\,7.6 & 1.78 & n & & & (9) \\
NGC\,5907 & 0.2788 & 14.9 & 11.7 & 1.64 & d & & & (9) \\
UGC\,2092 & 0.3281 & 82.1 & ~\,3.0 & 0.99 & pl & & & (7) \\
NGC\,5746 & 0.1256 & 22.9 & ~\,6.9 & 0.86 & d,n & & & (9) \\
NGC\,4565 & 0.2381 & ~\,9.7 & 14.8 & 0.68 & n & & & (8) \\
IC\,2233 & 0.3464 & ~\,7.7 & ~\,4.6 & 0.32 & pl & $\circ$ & & (9);(12) \\
NGC\,4244 & 0.1466 & ~\,3.1 & 14.8 & 0.07 & eh2 & $\circ$ & & (5);(13) \\
\hline
NGC\,4183 & & 17.0 & ~\,5.1 & & d & & & (1) \\
NGC\,5023 & & ~\,6.0 & ~\,5.9 & & pl? & & & (9) \\
UGC\,9242 & & 26.3 & ~\,4.9 & & ee? & & & (5) \\
\hline
\end{tabular}
\end{flushleft}
\hspace*{0.3cm}{\footnotesize $^{a}$ a$=$arc(s), d$=$disk emission (only 
planar DIG), ee$=$extended emission (locally), eh2$=$extraplanar 
\hbox{H\,{\sc ii}} region(s),\\ 
\hspace*{0.3cm} f$=$filament(s), h$_{\rm b}$=bright halo, h$_{\rm f}$=faint 
halo, n$=$no (e)DIG, pa$=$patch(es), pec$=$peculiar, pl$=$plume(s)\\}
\vspace*{0.2cm}
\hspace*{0.3cm}{\footnotesize References: (1) Collins et al.~(2000); (2) 
English \& Irwin (1997); (3) Ferguson et al.~(1996); (4) Hoopes et 
al.~(1996);\\ 
\hspace*{0.15cm} (5) Hoopes et al.~(1999); (6) Kenney \& Koopmann (1999); 
(7) Pildis et al.~(1994); (8) Rand et al.~(1992); (9) Rand (1996);\\
\hspace*{0.15cm} (10) Wang et al.~(2002); (11) Irwin et al.~(2000); (12) 
Hummel et al.~(1991); (13) Hummel et al.~(1984)\\ 
\hspace*{0.15cm} (14) Schlegel et al.~(1997) ($\bullet$ = detection, 
$\circ$ = non--detection)}
\end{table*}

\clearpage

\begin{table}
\setcounter{table}{3}
\caption[]{Starburst sample LH95 for comparison}
\begin{flushleft}
\begin{center}
\begin{tabular}{lcc}
\noalign{\smallskip}
\hline\hline
Galaxy & $S_{60}/S_{100}$ & $L_{\rm{FIR}}/D^2_{25}$ \\ 
 & & [$\rm{10^{40}\,erg\,s^{-1}\,kpc^{-2}}$] \\
\hline
IIZw035 & 0.8551 & 479.93 \\
IIZw096 & 1.1981 & 439.65 \\
IRAS\,03359--1523 & 0.8923 & 404.65 \\
IRAS\,09143+0939 & 0.8947 & 248.21 \\
NGC\,4418 & 1.3734 & 226.99 \\
IRAS\,05447--2114 & 0.8333 & 129.32 \\
IRAS\,12018+1941 & 0.8889 & 106.84 \\
NGC\,7552 & 0.7225 & \,~89.63 \\
NGC\,5104 & 0.5234 & \,~78.86 \\
Circinus & 0.7880 & \,~78.61 \\
ESO\,485--003 & 0.5289 & \,~77.59 \\
NGC\,6240 & 0.8958 & \,~73.75 \\
NGC\,5900 & 0.4431 & \,~70.62 \\
NGC\,2146 & 0.7609 & \,~66.37 \\
UGC\,6436 & 0.5714 & \,~62.49 \\
IC\,5179 & 0.5256 & \,~54.03 \\
NGC\,4433 & 0.6339 & \,~48.39 \\
ES0\,484--036 & 0.5400 & \,~47.93 \\
NGC\,2798 & 0.7038 & \,~44.27 \\
UGC\,903 & 0.5411 & \,~38.12 \\
NGC\,4666 & 0.4499 & \,~33.92 \\
NGC\,7541 & 0.5074 & \,~30.85 \\
NGC\,1572 & 0.4639 & \,~29.89 \\
NGC\,1511 & 0.5446 & \,~27.49 \\
NGC\,1808 & 0.7114 & \,~27.01 \\
NGC\,3885 & 0.7260 & \,~26.15 \\
NGC\,4818 & 0.7519 & \,~25.89 \\
NGC\,1134 & 0.5590 & \,~22.59 \\
NGC\,253 & 0.5005 & \,~21.79 \\
NGC\,5775$^\star$ & 0.3415 & \,~21.49 \\
NGC\,3034 & 1.1625 & \,~21.39 \\
NGC\,5253 & 1.0470 & \,~17.54 \\
NGC\,2966 & 0.6667 & \,~17.52 \\
NGC\,660 & 0.6887 & \,~15.81 \\
NGC\,3627 & 0.4934 & \,~15.38 \\
NGC\,4527 & 0.5053 & \,~15.19 \\
NGC\,4945$^\star$ & 0.5661 & \,~14.79\\
NGC\,4088 & 0.4375 & \,~13.21 \\
NGC\,3079 & 0.4855 & \,~12.73 \\
NGC\,5073 & 0.6791 & \,~11.64 \\
NGC\,3593 & 0.5083 & \,~10.81 \\
NGC\,4536 & 0.6435 & \,~\,~8.16 \\
NGC\,3556$^\dagger$ & 0.4164 & \,~\,~7.82 \\
NGC\,3044$^\star$ & 0.4633 & \,~\,~6.72 \\
NGC\,4388$^\star$ & 0.6127 & \,~\,~5.53 \\
NGC\,3628$^\star$ & 0.4694 & \,~\,~5.02 \\
NGC\,3067 & 0.4922 & \,~\,~5.00 \\
NGC\,3511 & 0.3805 & \,~\,~4.27 \\
NGC\,2820 & 0.4118 & \,~\,~3.81 \\
NGC\,3448 & 0.5299 & \,~\,~3.57 \\
\hline
\end{tabular}
\end{center}
\end{flushleft}
\end{table}


\begin{acknowledgements}
It is our sincere pleasure to express our thanks to Dr.~Francisco Prada for 
carrying out some of the observations at Calar Alto in an emergency case. 
We owe special thanks to Dr.~Michael Dahlem for providing us with the data 
on NGC\,3936, kindly observed by Dr.~Eva Grebel. We would also like to thank 
the anonymous referee for his/her comments which helped to improve the 
clarity of the text in a few places. The authors would like to thank 
Deutsches Zentrum f\"ur Luft-- und Raumfahrt (DLR) for financial support of 
this research project through grant 50\,OR\,9707. Additional travel support 
for the Calar Alto observing runs is acknowledged from the DFG through 
various grants. This research has made extensive use of the NASA/IPAC 
Extragalactic Database (NED) which is operated by the Jet Propulsion 
Laboratory, California Institute of Technology, under contract with the 
National Aeronautics and Space Administration.        
\end{acknowledgements}

\end{document}